\documentclass[aps,superscriptaddress,notitlepage,onecolumn]{revtex4-2}
\pdfoutput=1

\usepackage{amsthm,amsmath,amssymb,bm}
\usepackage{natbib}
\usepackage[utf8]{inputenc}
\usepackage{soul}
\usepackage{braket}
\usepackage{xcolor}
\usepackage{graphicx}

\definecolor{VibrantBlue}{HTML}{0077BB}
\definecolor{VibrantCyan}{HTML}{33BBEE}
\definecolor{VibrantTeal}{HTML}{009988}
\definecolor{VibrantOrange}{HTML}{EE7733}
\definecolor{VibrantRed}{HTML}{CC3311}
\definecolor{VibrantMagenta}{HTML}{EE3377}
\definecolor{VibrantGrey}{HTML}{BBBBBB}
\usepackage[colorlinks=true, urlcolor=VibrantBlue, citecolor=VibrantBlue,
linkcolor=VibrantMagenta, anchorcolor=VibrantMagenta]{hyperref}

\DeclareMathOperator*{\argmin}{arg\,min}
\DeclareMathOperator{\cvar}{\text{CVaR}}
\DeclareMathOperator{\real}{\text{Re}}

\begin{document}

\title{A case study of variational quantum algorithms for a job shop scheduling problem}

\author{David Amaro}
\email{david.amaro@cambridgequantum.com}
\affiliation{Cambridge Quantum Computing Limited, SW1P 1BX London, United Kingdom}

\author{Matthias Rosenkranz}
\email{matthias.rosenkranz@cambridgequantum.com}
\affiliation{Cambridge Quantum Computing Limited, SW1P 1BX London, United Kingdom}

\author{Nathan Fitzpatrick}
\affiliation{Cambridge Quantum Computing Limited, Terrington House 13-15, Hills Road, Cambridge, CB2 1NL, United Kingdom}

\author{Koji Hirano}
\affiliation{Nippon Steel Corporation, 20-1 Shintomi, Futtsu, Chiba, 293-8511, Japan}

\author{Mattia Fiorentini}
\affiliation{Cambridge Quantum Computing Limited, SW1P 1BX London, United Kingdom}

\date{\today}

\begin{abstract}
Combinatorial optimization models a vast range of industrial processes aiming at
improving their efficiency. In general, solving this type of problem exactly is
computationally intractable. Therefore, practitioners rely on heuristic solution
approaches. Variational quantum algorithms are optimization heuristics that can
be demonstrated with available quantum hardware. In this case study, we apply
four variational quantum heuristics running on IBM's superconducting quantum
processors to the job shop scheduling problem. Our problem optimizes a steel
manufacturing process. A comparison on 5 qubits shows that the recent filtering
variational quantum eigensolver (F-VQE) converges faster and samples the global
optimum more frequently than the quantum approximate optimization algorithm
(QAOA), the standard variational quantum eigensolver (VQE), and variational
quantum imaginary time evolution (VarQITE). Furthermore,  F-VQE readily solves
problem sizes of up to 23 qubits on hardware without error mitigation post
processing.
\end{abstract}

\maketitle

\section{Introduction}
One of the major drivers of industry's recent interest in quantum computing is
the promise of improving combinatorial optimization. This could have a large
impact across many sectors including manufacturing, finance, logistics and
supply chain management. However, most combinatorial optimization problems are
NP-hard making it unlikely that even quantum computers can solve them
efficiently in the worst-case. Informally, NP-hardness means that finding exact
solutions is not more efficient than going through all potential solutions---at
a cost that grows exponentially with the problem size. Quantum algorithms such
as Grover's perform exhaustive search with a quadratic speedup but require fault
tolerant quantum
hardware~\cite{groverFastQuantumMechanical1996a,durrQuantumAlgorithmFinding1999}.
Instead it is interesting to explore if quantum computers could speed up the
average-case or special cases of practical interest or, indeed, improve
approximate solutions in practice on non-fault-tolerant hardware.

A large body of research focuses on quantum-enhanced optimization heuristics for
the noisy intermediate-scale quantum (NISQ)
era~\cite{Preskill2018,cerezoVariationalQuantumAlgorithms2021,bhartiNoisyIntermediatescaleQuantum2021}.
Typically, these algorithms don't come equipped with convergence guarantees and
instead solve the problem approximately within a given computational budget.
While many fault-tolerant optimization algorithms can also be formulated as
heuristics~\cite{sandersCompilationFaultTolerantQuantum2020}, our focus is on
variational quantum algorithms (VQA). Typically VQA employ objective functions
implemented with parameterized quantum circuits (PQCs) and update their
parameters via a classical optimization routine. In our context, a common
approach for combinatorial optimization encodes the optimal solution in the
ground state of a classical multi-qubit Hamiltonian~\cite{Kochenberger2014,
Lu14, Glover2019}.

Studying the effectiveness of such heuristics relies on intuition and
experimentation. However, today's quantum computers are noisy and fairly limited
in size making such experimentation hard. Nevertheless it is important to gauge
properties such as convergence speed, scalability and accuracy from the limited
hardware we have available. To make the most of today's NISQ computers it is
reasonable to compare different VQA on concrete problems.

We selected the popular quantum approximate optimization algorithm
(QAOA)~\cite{Farhi2014} and the variational quantum eigensolver
(VQE)~\cite{Peruzzo2014} as well as the less well studied variational quantum
imaginary time evolution algorithm (VarQITE)~\cite{Mcardle2019} and the
filtering variational quantum eigensolver
(F-VQE)~\cite{amaroFilteringVariationalQuantum2022} recently introduced by some
of the present authors. Despite its promising properties, such as supporting a
form of quantum advantage,~\cite{Farhi2016, Zhou2020, Moussa2020} and
considerable progress with regards to its experimental
realization~\cite{harriganQuantumApproximateOptimization2021}, in general the
QAOA ansatz requires circuit depths that are challenging for current quantum
hardware. VQE, VarQITE and F-VQE employ more flexible, hardware-efficient
ans\"atze tailored for the particular quantum processor. Those ans\"atze feature
high expressibility and  entangling
capabilities~\cite{simExpressibilityEntanglingCapability2019a}, which suggests
that they can lead to genuinely different heuristics compared to classical ones.
On the other hand, they are prone to barren plateaus which could prevent the
algorithms' convergence at larger problem
sizes~\cite{mccleanBarrenPlateausQuantum2018,cerezoCostFunctionDependent2021}.
In addition, the classical optimizer can significantly affect the performance of
quantum heuristics on NISQ hardware, and the magnitude of this effect can vary
between optimization
problems~\cite{guerreschiPracticalOptimizationHybrid2017,nanniciniPerformanceHybridQuantumclassical2019,lavrijsenClassicalOptimizersNoisy2020,sungUsingModelsImprove2020,pellow-jarmanComparisonVariousClassical2021}.
Those effects have made it difficult in the past to scale common VQA beyond
small-scale experiments. Here we compare VQA executed on IBM's superconducting
quantum computers with a view towards scaling up a particular optimization
problem of industrial relevance.

We compare the effectiveness of VQE, QAOA, VarQITE and F-VQE on the job shop
scheduling problem (JSP). The JSP is a combinatorial optimization problem where
jobs are assigned to time slots in a number of machines or processes in order to
produce a final product at minimal cost. Typically costs are associated with
delivery delays or reconfiguration of production processes between time slots.
The JSP formulation considered herein was developed by Nippon Steel Corporation
and applies to processes typical of steel manufacturing.

This article is structured as follows. Sec.~\ref{sec:methods} introduces the JSP
formulation and the four VQA employed in this work, highlighting their
similarities and differences. Sec.~\ref{sec:results} analyses the performance of
all VQA and shows results of scaling up F-VQE on hardware. We conclude in
Sec.~\ref{sec:conclusions}. Appendix~\ref{sec:derivation} includes a derivation
of the JSP formulation, App.~\ref{sec:jsp-scaling} discusses the scaling of the
JSP, App.~\ref{sec:hardware} lists key properties of the quantum processors used
for this work, and App.~\ref{sec:additional} provides several additional results
from hardware experiments.

\section{Methods}\label{sec:methods}

This section introduces the JSP and its mathematical formulation in
Secs. \ref{ssec:jsp}--\ref{ssec:qubo} and introduces the VQE, QAOA, VarQITE and F-VQE with our
choices for the various settings of these algorithms in
Sec.~\ref{ssec:algorithms}.

\subsection{Job shop scheduling in a steel manufacturing
process}\label{ssec:jsp} The general JSP is the problem of finding an
assignment---also called a schedule--- of $J$ jobs to $M$ machines, where each
job needs to be processed in a certain order across the machines. Each job can
carry additional data such as due time or processing time. A JSP is typically
described by two further components: \emph{processing characteristics and
constraints} and an \emph{objective}. The processing characteristics and
constraints encode the specifics of an application such as setup times of
machines and job families or production groups. Typical examples of objectives
to minimise include makespan (total completion time) or mismatch of the jobs'
completion and due times (for an overview of common scheduling formulations, see
Ref.~\cite{pinedoSchedulingTheoryAlgorithms2012}).

The JSP formulation we consider applies to general manufacturing processes and
was fine-tuned by Nippon Steel Corporation for steel manufacturing.  We consider
jobs $j=1, \dots, J$ assigned to different machines or processes $m=1, \dots, M$
at time slots $t_m =1, \dots, T_m$. In this work, the processing times of all
jobs for all processes are assumed to be equal. Accordingly, time slots can be
common across the multiple processes and thus $t_m$ is simplified as $t$
throughout the paper. The processing times of all jobs are equal and each job is
assigned a due time $d_j$. Each machine $m$ is allowed to idle for a total
number of time slots $e_m \geq 0$ at the beginning or end of the schedule. This
number is an input of the problem. Hence, the maximum time slot for machine $m$
is $T_m = J + e_m$.

The objective is to minimize the sum of \emph{early delivery} and \emph{late
delivery} of jobs leaving the last machine, and the \emph{production cost}
associated with changing the processing conditions for subsequent jobs in each
machine. Early (late) delivery is quantified by a constant $c_e$ ($c_l$)
multiplied by the number of time steps a job finishes before (after) its due
date, summed over all jobs. To compute the production cost for each machine $m$
each job $j$ is assigned a \emph{production group} $P_{mj}$. The production cost
is quantified by a constant $c_p$ multiplied by the total number of times
consecutive jobs $j_1$, $j_2$ in a machine $m$ switch productions groups, i.e.
$P_{mj_1} \neq P_{mj_2}$. Figure~\ref{fig:instance} illustrates these costs for
the largest (20-job) JSP instance we consider in this work.

We consider the following sets of constraints, which follow from the specifics of the manufacturing process.
\begin{enumerate}
    \item \emph{Job assignment constraints.} Each job is assigned to exactly one time slot in each machine.
    \item \emph{Time assignment constraints.} $J$ jobs are distributed to $J$ consecutive time slots in each machine.
    \item \emph{Process order constraints.} Each job must progress from machine $1$ to $M$ in non-descending order.
    \item \emph{Idle slot constraints.} Idle slots occur only at the beginning or end of a schedule.
\end{enumerate}

\subsection{Quadratic unconstrained binary optimization formulation of the
JSP}\label{ssec:qubo}

We formulate the JSP defined in subsection~\ref{ssec:jsp} as a Quadratic
Unconstrained Binary Optimization (QUBO) problem. A feasible solution of the JSP
is a set of two schedules $(\bm{x}, \bm{y})$ given by binary vectors $\bm{x} \in
\mathbb{B}^{N_x}$ for the \emph{real jobs} (those corresponding to jobs $1,
\dots, J$) and $\bm{y} \in \mathbb{B}^{N_y}$ for the \emph{dummy jobs}
introduced to fill idle time slots at the beginning and end of each machine's
schedule. Here $\mathbb{B} = \set{0, 1}$, $N_x = \sum_{m = 1}^M{J (J + e_m)}$,
and $N_y = \sum_{m = 1}^M {e_m}$. $N_y$ is independent of $J$ because, owing to
the idle slot constraints, the optimization only needs to decide on the number
of consecutive dummy jobs at the beginning of the schedule per machine. A value
$x_{mjt}=1$ ($x_{mjt}=0$) indicates that job $j$ is assigned (is not assigned)
to machine $m$ at time $t$. Similarly, for dummy jobs, value $y_{mt}=1$
($y_{mt}=0$) indicates that a dummy job is (is not) assigned to machine $m$ at
time slot $t$. With the cost and constraints of the JSP  encoded in a
\emph{quadratic form} $Q\colon \mathbb{B}^{N_x} \times \mathbb{B}^{N_y}
\rightarrow \mathbb{R}$ the JSP becomes
\begin{equation}\label{eq:JSP}
    (\bm{x}^*, \bm{y}^*) = \argmin_{(\bm{x}, \bm{y}) \in \mathbb{B}^{N_x} \times \mathbb{B}^{N_y}} Q(\bm{x}, \bm{y}).
\end{equation}
The binary representation makes it straightforward to embed the problem on a
quantum computer by mapping schedules to qubits.

\begin{figure}[t]
    \centering
    \includegraphics{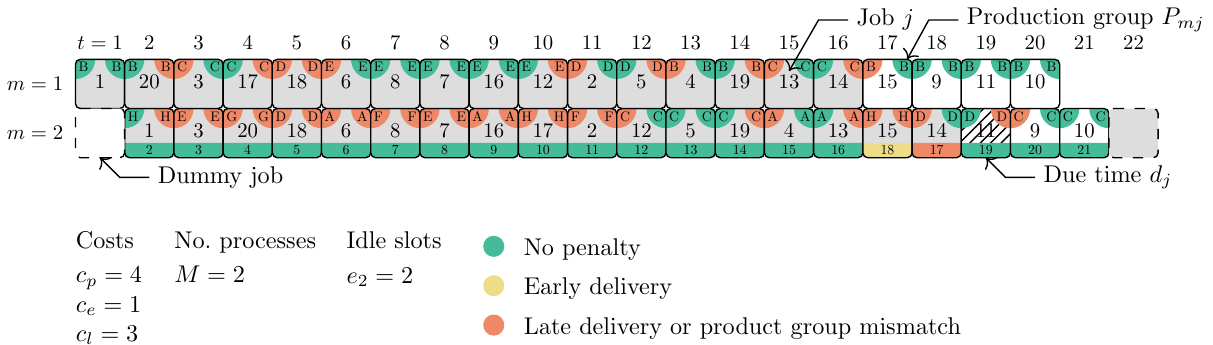}
    \caption{20-job, 2-process JSP instance considered in this work and its
    optimal solutions. Colors at the bottom of each box indicate whether early
    or late delivery costs apply for each time slot. Colors in the corners of
    each box indicate whether the production cost applies for each machine and
    consecutive time slot. By fixing some jobs to their optimal slots we
    generate instances with different numbers of free variables $N$. This is
    indicated by the background color/pattern of a box: grey for fixed slots and
    jobs, white for free slots and jobs, and dashes for free slots but fixed
    jobs. We generated instances with $N=5, 10, 12, 16, 23$ free variables (see
    Tab.~\ref{tab:prob_inst}). The figure shows $N=23$.}
    \label{fig:instance}
\end{figure}

\begin{table}[t]
    \begin{tabular}{cccc}
        \hline
        Free variables $N$ & Machine $m$ & Time slots & Jobs \\
        \hline
        5 & 2 & 1, 20 - 21 & 9 - 10 \\
        10 & 2 & 1, 19 - 21 & 9 - 11 \\
        12 & 2 & 19 - 22 & 9 - 11 \\
        16 & 2 & 18 - 21 & 9 - 11, 14 \\
        23 & 1; 2 &  17 - 20; 1, 19 - 21    & 9 - 11, 15; 9 - 10 \\
        \hline
    \end{tabular}
    \caption{Time slots and jobs needing assignment in each of the problem instances considered in this work.}
    \label{tab:prob_inst}
\end{table}

The function $Q$ for the JSP is
\begin{equation} \label{eq:qubo}
	\begin{split}
		Q(\bm{x}, \bm{y}) &= c(\bm{x}) + p \sum_{m=1}^{M} \sum_{j=1}^{J} (g_{mj}(\bm{x}) - 1)^2 + p \sum_{m=1}^{M} \sum_{t=1}^{T_m} (\ell_{mt}(\bm{x}, \bm{y}) - 1)^2 \\
		&\quad + p \sum_{m=1}^{M-1} \sum_{j=1}^{J} q_{mj}(\bm{x}) + p \sum_{m=2}^{M} \sum_{t=1}^{e_m-1} r_{mt}(\bm{y}).
	\end{split}
\end{equation}
All terms are derived in more detail in App.~\ref{sec:derivation}. $c(\bm{x})$
is the cost of the schedule, Eq.~\eqref{eq:cost}, $g_{mj}(\bm{x})$ encodes the
job assignment constraints, Eq.~\eqref{eq:real_job}, $\ell_{mt}(\bm{x}, \bm{y})$
encodes the time assignment constraints, Eq.~\eqref{eq:time}, $q_{mj}(\bm{x})$
encodes the process order constraints, Eq.~\eqref{eq:process_order},
$r_{mt}(\bm{x})$ encodes the idle slot constraints,
Eq.~\eqref{eq:idle_slot_cons}. The constraints are multiplied by a penalty $p$,
which will be set to a sufficiently large value. To ensure non-negative
penalties some constraints need to be squared. Note that $Q$ is a quadratic form
because all terms can be written as polynomials of degree two in the binary
variables $\bm{x}$ and $\bm{y}$. To simplify notation we often denote the
concatenation of the two sets of binary variables with $\bm{z} = (\bm{x},
\bm{y})$ and $Q(\bm{z}) = Q(\bm{x}, \bm{y})$.  Fig.~\ref{fig:instance}
illustrates the largest JSP instance used in this work together with its optimal
solution obtained via a classical solver, and Tab.~\ref{tab:prob_inst} specifies
all instances used. App.~\ref{sec:jsp-scaling} derives the scaling of the total
number of variables for this formulation.

Solving the JSP, Eq.~\eqref{eq:JSP}, is equivalent to finding the ground state
of the Hamiltonian
\begin{equation}\label{eq:H}
    H = Q\left( \frac{I - \bm{Z}^{(\bm{x})}}{2}, \frac{I - \bm{Z}^{(\bm{y})}}{2}\right) = h_0 I + \sum_{n=1}^N h_n Z_n + \sum_{n,n'=1}^N h_{nn'} Z_n Z_{n'}
\end{equation}
where the vectors of Pauli $Z$ operators $\bm{Z}^{(\bm{x})}, \bm{Z}^{(\bm{y})}$ correspond
to the binary variables in $\bm{x}, \bm{y}$, respectively, $\bm{Z}$
corresponds to $\bm{z}$, and $h_0$, $h_n$, $h_{nn'}$ are the coefficients of the
corresponding operators. Note that this Hamiltonian is defined purely in terms
of Pauli $Z$ operators, which means that its eigenstates are separable and
they are computational basis states.

\subsection{Variational quantum algorithms for combinatorial optimization problems}\label{ssec:algorithms}

\begin{table}[t]
    \centering
    \begin{tabular}{p{0.1\linewidth}p{0.023\linewidth}p{0.17\linewidth}p{0.19\linewidth}p{0.2\linewidth}p{0.16\linewidth}}
        \hline
        & $N$ & \textbf{VQE} & \textbf{QAOA} & \textbf{VarQITE} & \textbf{F-VQE}\\
        \hline
        \textbf{Ansatz} & 5\newline\hspace*{-2.2ex} $>$5& Fig.~\ref{fig:ansatz} ($p=2$)\newline -- & Eq.~\ref{eq:PQC-QAOA} ($p=2$)\newline -- & Fig.~\ref{fig:ansatz} ($p=2$)\newline -- & Fig.~\ref{fig:ansatz} ($p=2$)\newline Fig.~\ref{fig:ansatz} ($p=1$)\\
        \textbf{Initial param.} & & $\ket{+}^{\otimes N}$ & uniform in $[0, \pi]$ & $\ket{+}^{\otimes N}$ & $\ket{+}^{\otimes N}$\\
        \textbf{Objective} & & CVaR Eq.~\eqref{eq:objective-VQE}\newline
        ($\alpha=0.5$) & CVaR Eq.~\eqref{eq:objective-QAOA}\newline
        ($\alpha=0.5$) & Mean energy Eq.~\eqref{eq:objective-QITE} & Custom
        Eq.~\eqref{eq:objective-F-VQE}\\
        \textbf{Optimizer} & & COBYLA & COBYLA & Eq.~\eqref{eq:optimizer-QITE} & Eq.~\eqref{eq:F-VQE_update}\\
        \textbf{No. shots} & $5$\newline $10$\newline $12$\newline $16$\newline $23$ & 1,000\newline --\newline --\newline --\newline -- & 1,000\newline --\newline --\newline --\newline -- & 1,000\newline --\newline --\newline --\newline -- & 1,000 \newline 500\newline 550\newline 650 \newline 450\\
        \textbf{Quantum chip} & $5$\newline $10$\newline $12$\newline $16$\newline $23$ & multiple\newline --\newline --\newline --\newline -- & multiple\newline --\newline --\newline --\newline -- & multiple\newline --\newline --\newline --\newline -- & multiple\newline ibmq\_toronto\newline ibmq\_guadalupe\newline ibmq\_manhattan\newline ibmq\_manhattan \\
        \textbf{Key\newline findings} & & \raggedright Flexible ansatz;\newline
        converges slower than F-VQE
        & \raggedright Ansatz fixed by problem topology;\newline
        poor convergence likely due to noise & \raggedright Flexible ansatz;\newline strongly varying performance across runs;
        converges slower than F-VQE & \raggedright \textbf{Flexible
        ansatz;\newline fastest, most consistent convergence} \\
    \end{tabular}
    \caption{VQA and settings used for the hardware experiments in
    Figs.~\ref{fig:comparison_vqe_qaoa}-\ref{fig:comparison_fvqe}. An initial
    parameter $\ket{+}^{\otimes N}$ means that the initial angles of all $R_y$
    in the first (second) layer of the ansatz are set to $0$ ($\pi/2$). The last
    line highlights some key findings from our experiments.}
    \label{tab:VQA}
\end{table}

VQA are the predominant paradigm for algorithm development on gate-based NISQ
computers. They comprise several components that can be combined and adapted in
many ways making them very flexible for the rapidly changing landscape of
quantum hard- and software development. The main components are an \emph{ansatz}
for a PQC, a \emph{measurement scheme}, an \emph{objective function}, and a
classical \emph{optimizer}. The measurement scheme specifies the operators to be
measured, the objective function combines measurement results in a classical
function, and the optimizer proposes parameter updates for the PQC with the goal
of minimising the objective function. As noted in subsection~\ref{ssec:qubo},
the JSP is equivalent to finding the ground state of the Hamiltonian
Eq.~\eqref{eq:H}. VQA are well suited to perform this search by minimising a
suitable objective function. We focus on four VQA for solving the JSP: VQE,
QAOA, VarQITE, and F-VQE.

We use conditional Value-at-Risk (CVaR) as the objective function for all
VQA~\cite{Barkoutsos2020}. For a random variable $X$ with quantile function
$F^{-1}$ the CVaR is defined as the conditional expectation over the left tail
of the distribution of $X$ up to a quantile $\alpha \in (0, 1]$:
\begin{equation}\label{eq:cvar}
    \cvar_\alpha(X) = \mathbb{E}[X | X \leq F_X^{-1}(\alpha)].
\end{equation}
In practice we estimate the CVaR from measurement samples as follows. Prepare a
state $\ket{\psi}$ and measure this state $K$ times in the computational basis.
Each measurement corresponds to a bitstring $\bm{z}_k$ sampled from the
distribution implied by the state $\ket{\psi}$ via the Born rule, $\bm{z}_k \sim
\vert \braket{\bm{z} | \psi} \vert^2$.  We interpret each bitstring as a
potential solution to the JSP with energy (or cost) $E_k = Q(\bm{z}_k)$, $k=1,
\dots, K$. Given a sample of energies $\set{E_1, \dots, E_K}$---without loss of
generality assumed to be ordered from small to large---the CVaR estimator is
\begin{equation}\label{eq:cvar-estimator}
    \widehat{\cvar}_\alpha(\set{E_1, \dots, E_K}) = \frac{1}{\lceil \alpha K\rceil}  \sum_{k=1}^{\lceil \alpha K \rceil} E_k.
\end{equation}
For $\alpha=1$ the CVaR estimator is the sample mean of energies, which is the
objective function often used in standard VQE. The CVaR estimator with $0 <
\alpha < 1$ has shown advantages in applications that aim at finding ground
states, such as combinatorial optimization problems~\cite{Barkoutsos2020} and
some of our experiments confirmed this behaviour.

The difference between the considered VQA boils down to different choices of the
ansatz, measurement scheme, objective and optimizer. Table~\ref{tab:VQA}
compares the four algorithms and our concrete settings and
subsections~\ref{sssec:VQE}--\ref{sssec:F-VQE} detail the algorithms.
Appendix~\ref{sec:hardware} lists the quantum processors used for the hardware
execution.

\subsubsection{Variational Quantum Eigensolver}\label{sssec:VQE}
\begin{figure}
    \centering
    \includegraphics{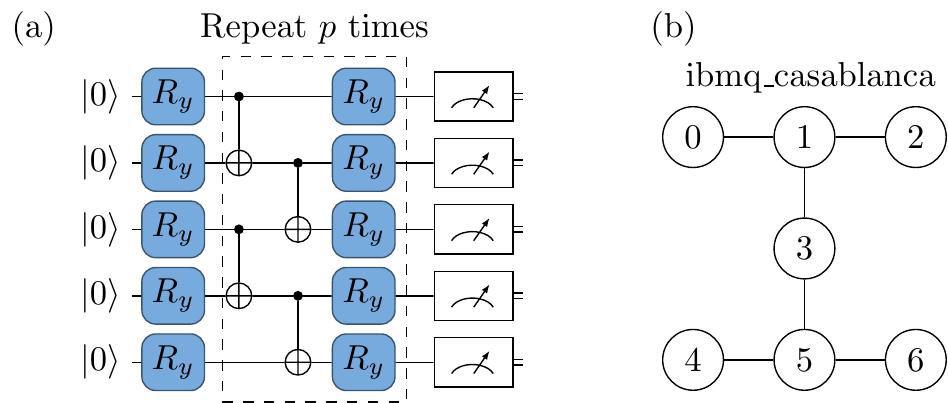}
    \caption{(a) Parameterized quantum circuit ansatz $\ket{\psi(\bm{\theta})}$
    and (b) connectivity of the ibmq\_casablanca quantum processor used for the
    5-qubit VQE, VarQITE and F-VQE results. Each $R_y$ in (a) is a single-qubit
    rotation gate rotating the qubit around the $Y$ axis by an individual angle
    $\theta$ per gate, $R_y=R_y(\theta)=\exp(-i\theta Y/2)$. Gates in the dashed
    box are repeated $p$ times, where $p$ is the number of layers. In (b) each
    circle is a physical qubit and lines indicate their physical connectivity.}
    \label{fig:ansatz}
\end{figure}

VQE aims at finding the lowest energy state within a family of parameterized
quantum states. It was introduced for estimating the ground state energies of
molecules described by a Hamiltonian in the context of quantum chemistry.
Exactly describing molecular ground states would require an exponential number
of parameters. VQE offers a way to approximate their description using a
polynomial number of parameters in a PQC ansatz. Since the JSP can be expressed
as the problem of finding a ground state of the Hamiltonian Eq.~\eqref{eq:H},
VQE can also be used for solving the JSP. This results in a heuristic
optimization algorithm for the JSP similar in spirit to classical heuristics,
which aim at finding good approximate solutions.

Our VQE implementation employs the hardware-efficient ansatz in
Fig~\ref{fig:ansatz}(a) for the PQC. Hardware-efficient ans\"atze are very
flexible as they can be optimized for a native gate set and topology of a given
quantum processor~\cite{Kandala2017}. We denote the free parameters of the
single-qubit rotation gates in the ansatz with the vector $\bm{\theta}$. The PQC
implements the unitary operator $U(\bm{\theta})$ and $\ket{\psi({\bm{\theta}})}
= U(\bm{\theta}) \ket{0}$ denotes the parameterized state after executing this
PQC.

The measurement scheme for VQE is determined by the Hamiltonian we wish to
minimize. In the case of JSP this reduces to measuring tensor products of Pauli
$Z$ operators given by Eq.~\eqref{eq:H}. All terms commute so they can be
computed from a single classical bitstring $\bm{z}_k \sim \vert \braket{\bm{z} |
\psi(\bm{\theta})} \vert^2$ sampled from the PQC. Sampling $K$ bitstrings and
calculating their energies $E_k(\bm{\theta}) = Q(\bm{z}_k(\bm{\theta}))$ yields
a sample of (ordered) energies $\set{E_1(\bm{\theta}), \dots, E_K(\bm{\theta})}$
parameterized by $\bm{\theta}$. Plugging this sample into the CVaR estimator,
Eq.~\eqref{eq:cvar}, yields the objective function for VQE
\begin{equation}\label{eq:objective-VQE}
    O_\text{VQE}(\bm{\theta}; \alpha) = \widehat{\cvar}_\alpha(\set{E_1(\bm{\theta}), \dots, E_K(\bm{\theta})}).
\end{equation}

We use the Constrained Optimization By Linear Approximation (COBYLA) optimizer
to tune the parameters of the PQC~\cite{powellDirectSearchOptimization1994}.
This is a gradient-free optimizer with few hyperparameters making it a
reasonable baseline choice for VQA~\cite{lavrijsenClassicalOptimizersNoisy2020}.

\subsubsection{Quantum Approximate Optimization Algorithm}\label{sssec:QAOA}

QAOA is a VQA which aims at finding approximate solutions to combinatorial
optimization problems. In contrast to VQE, research on QAOA strongly focuses on
combinatorial optimization rather than chemistry problems. QAOA can be thought
of as a discretized approximation to quantum adiabatic
computation~\cite{farhiQuantumComputationAdiabatic2000}.

The QAOA ansatz follows from applying the two unitary operators $U_M(\beta) =
e^{-i\beta \sum_{n=1}^N X_n}$ and $U(\gamma) = e^{-i\gamma H}$ a number of $p$
times to the $N$-qubit uniform superposition $\ket{+} = \frac{1}{\sqrt{2^N}}
\sum_{n=0}^{2^N-1} \ket{n}$ in an alternating sequence. Here $X_n$ is the Pauli
$X$ operator applied to qubit $n$ and $H$ is the JSP Hamiltonian,
Eq.~\eqref{eq:H}. The QAOA ansatz with $2p$ parameters $(\bm{\beta},
\bm{\gamma})$ is
\begin{equation}\label{eq:PQC-QAOA}
    \ket{\psi(\bm{\beta}, \bm{\gamma})} = U_M(\beta_p)U(\gamma_p) U_M(\beta_{p-1})U(\gamma_{p-1}) \cdots U_M(\beta_1)U(\gamma_1) \ket{+}.
\end{equation}
In contrast to our ansatz for VQE, in the QAOA ansatz the connectivity of the
JSP Hamiltonian dictates the connectivity of the two-qubit gates. This means
that implementing this ansatz on digital quantum processors with physical
connectivity different from the JSP connectivity requires the introduction of
additional gates for routing. This overhead can be partly compensated by clever
circuit optimization during the compilation stage.

We use the same measurement scheme, objective function and optimizer for QAOA
and VQE. Namely, we sample bitstrings $\bm{z}_k(\bm{\beta}, \bm{\gamma})$ from
the PQC and calculate their energies $E_k(\bm{\beta}, \bm{\gamma}) =
Q(\bm{z}_k(\bm{\beta}, \bm{\gamma}))$. The objective function is the CVaR
estimator
\begin{equation}\label{eq:objective-QAOA}
    O_\text{QAOA}(\bm{\beta}, \bm{\gamma}; \alpha) = \widehat{\cvar}_\alpha(\set{E_1(\bm{\beta}, \bm{\gamma}), \dots, E_K(\bm{\beta}, \bm{\gamma})})
\end{equation}
and the optimizer is COBYLA.

\subsubsection{Variational Quantum Imaginary Time
Evolution}\label{sssec:VarQITE}

Imaginary time evolution is a technique for finding ground states by evolving an
initial state with the Schr\"odinger equation in imaginary time $\tau = it$.
This technique has mainly been applied to study quantum many-body
problems~\cite{yuanTheoryVariationalQuantum2019} and a variant of the algorithm
shows promising results for combinatorial optimization~\cite{Motta2020}. Here we
use a variational formulation of imaginary time evolution dubbed
VarQITE~\cite{Mcardle2019} to find approximate solutions of the JSP.

Given an initial state $\ket{\phi(0)}$ the imaginary time evolution is defined
by $\ket{\phi(\tau)} = e^{-H\tau} \ket{\phi(0)} / \sqrt{\mathcal{Z}(\tau)}$ with
a normalization factor $\mathcal{Z}(\tau) = \braket{\phi(0)\lvert e^{-2H\tau}
\rvert \phi(0)}$. The non-unitary operator $e^{-H\tau}$ cannot be mapped
directly to a quantum circuit and is typically implemented via additional qubits
and post-selection. To avoid additional qubits and post-selection, instead the
VarQITE algorithm optimizes a PQC to approximate the action of the non-unitary
operator. This is achieved by replacing the state $\ket{\phi(\tau)}$ with a
state $\ket{\psi(\bm{\theta})} = \ket{\psi(\bm{\theta}(\tau))} =
U(\bm{\theta})\ket{+}$ and the parameters are assumed to be time-dependent
$\bm{\theta} = \bm{\theta}(\tau)$. We use the PQC ansatz in
Fig.~\ref{fig:ansatz}(a) and set initial parameters such that the resulting
initial state is $\ket{+}$.

We use the same measurement scheme as in VQE with the mean energy as the
objective function, i.e. CVaR with $\alpha=1$,
\begin{equation}\label{eq:objective-QITE}
    O_\text{VarQITE}(\bm{\theta}) = \frac{1}{2}\widehat{\cvar}_1(\set{E_1(\bm{\theta}), \dots, E_K(\bm{\theta})}).
\end{equation}

VarQITE updates parameters with a gradient-based optimization scheme derived
from McLachlan's variational principle~\cite{yuanTheoryVariationalQuantum2019}.
This lifts the imaginary time evolution of the state $\ket{\phi(\tau)}$ to an
evolution of the parameters in the PQC via the differential equations
\begin{equation}\label{eq:varqite_evolution}
    A(\bm{\theta}) \frac{\partial\bm{\theta}(\tau)}{\partial\tau} = -\bm{\nabla}O_\text{VarQITE}(\bm{\theta}),
\end{equation}
where $A(\bm{\theta})$ is a matrix with entries
\begin{align}
    A_{ij} &= \real \left( \Braket{\frac{\partial\psi(\bm{\theta})}{\partial \theta_i} | \frac{\partial\psi(\bm{\theta})}{\partial \theta_j}} \right).
\end{align}
We assume small time steps $\delta\tau$, denote $\tau_n = \tau_n + n\delta\tau$,
$\bm{\theta}_{n} = \bm{\theta}(\tau_n)$ and approximate the parameter evolution
Eq.~\eqref{eq:varqite_evolution} with the explicit Euler scheme
\begin{equation}\label{eq:optimizer-QITE}
    \bm{\theta}_{n+1} = \bm{\theta}_{n} - A^{-1}\left(\bm{\theta}_{n}\right) \bm{\nabla}O_\text{VarQITE}(\bm{\theta}_n) \delta\tau.
\end{equation}
We estimate the entries of $A$ and $\bm{\nabla}O_\text{VarQITE}$ with the
Hadamard test. This requires an additional qubit and controlled operations.

\subsubsection{Filtering Variational Quantum Eigensolver}\label{sssec:F-VQE}

F-VQE is a generalization of VQE with faster and more reliable convergence to
the optimal solution~\cite{amaroFilteringVariationalQuantum2022}. The algorithm
uses \emph{filtering operators} to modify the energy landscape at each
optimization step. A filtering operator $f(H; \tau)$ for $\tau>0$ is defined via
a real-valued function $f(E; \tau)$ with the property that $f^2(E; \tau)$ is
strictly decreasing on the spectrum of the Hamiltonian $E \in [E_\text{min},
E_\text{max}]$.

For F-VQE we used the ansatz in Fig.~\ref{fig:ansatz}(a). In
contrast to our VQE implementation, F-VQE uses a gradient-based optimizer. At
each optimization step $n$ the objective function is
\begin{equation}\label{eq:objective-F-VQE}
    O_\text{F-VQE}^{(n)}(\bm{\theta}; \tau) = \frac{1}{2} \lVert \ket{\psi(\bm{\theta})} - \ket{F_n \psi_{n-1}} \rVert^2,
\end{equation}
where $\ket{\psi_{n-1}} = \ket{\psi(\bm{\theta}_{n-1})}$ and $\ket{F_n \psi_{n-1}} = F_n \ket{\psi_{n-1}} / \sqrt{\braket{F_n^2}_{\psi_{n-1}}}$ with $F_n = f(H; \tau_n)$. We use the \emph{inverse filter} $f(H; \tau) = H^{-\tau}$. It can be shown that the algorithm minimises the mean energy of the system, i.e. CVaR with $\alpha=1$. The update rule of the optimizer at step $n$ is
\begin{equation}\label{eq:F-VQE_update}
    \bm{\theta}_{n+1} = \bm{\theta}_{n} -  \eta\bm{\nabla}O_\text{F-VQE}^{(n)}(\bm{\theta}_n; \tau),
\end{equation}
where $\eta$ is a learning rate. The gradient in Eq.~\eqref{eq:F-VQE_update} is
computed with the parameter shift
rule~\cite{schuldEvaluatingAnalyticGradients2019,mitaraiQuantumCircuitLearning2018}.
This leads to terms of the form $\braket{F}_{\psi}$ and $\braket{F^2}_{\psi}$
for states $\ket{\psi}$. They can be estimated from bitstrings
$\bm{z}_k^\psi(\bm{\theta}) \sim \vert \braket{\bm{z} | \psi(\bm{\theta})}
\vert^2$ sampled from the PQC. A sample of $K$ bitstrings yields a sample of
filtered energies $\set{f_1^\psi(\bm{\theta}; \tau), \dots,
f_K^\psi(\bm{\theta}; \tau)}$ with $f_k^\psi(\bm{\theta}; \tau) =
f(Q(\bm{z}_k^\psi(\bm{\theta}); \tau)$. Then all $\braket{F}_\psi$ are estimated
from such samples via
\begin{equation}
    \braket{F}_\psi(\bm{\theta}; \tau) \approx \widehat{\cvar}_1(\set{f_1^\psi(\bm{\theta}; \tau), \dots, f_K^\psi(\bm{\theta}; \tau)})
\end{equation}
and equivalently for $\braket{F^2}_\psi$. Our implementation of F-VQE adapts the
parameter $\tau$ dynamically at each optimization step to keep the gradient norm
of the objective close to some large, fixed value (see
\cite{amaroFilteringVariationalQuantum2022} for details). 

\section{Results and Discussion}\label{sec:results}

We have tested the algorithms in Sec.~\ref{ssec:algorithms} on instances of the
JSP on IBM quantum processors. First we compared all algorithms on a 5-qubit
instance to evaluate their convergence. Then, based on its fast convergence, we
selected F-VQE to study the scaling to larger problem sizes. A comparison
against classical solvers is not in scope of this work (in fact, all instances
can be easily solved exactly). Instead we focus on convergence and scaling the
VQA for this particular optimization problem of industrial relevance. All
quantum processors were accessed via tket~\cite{sivarajahVertKetRangle2020}.
Hardware experiments benefitted from tket's out-of-the-box noise-aware qubit
placement and routing, but we did not use any other error mitigation techniques
involving additional post-processing.

All problem instances for the experiments have been obtained as sub-schedules of
the 20-job 2-machine problem whose solution is illustrated in Fig.
\ref{fig:instance}. Table \ref{tab:prob_inst} provides information on which
machine, time slot and job needed to be assigned a schedule in each of the
problem instances.

Throughout this section we plot average energies scaled to the range $[0, 1]$:
\begin{equation}\label{eq:eps_psi}
    \epsilon_{\psi} = \frac{\braket{H}_{\psi} - E_\text{min}}{E_\text{max} - E_\text{min}} \in [0, 1],
\end{equation}
where $E_\text{min}$, $E_\text{max}$ are the minimum and maximum energy of the
Hamiltonian, respectively, and $\braket{H}_{\psi}=\braket{\psi|H|\psi}$ for a
given state $\ket{\psi}$. We calculated $E_\text{min}$, $E_\text{max}$ exactly.
A value $\epsilon_{\psi} = 0$ corresponds to the optimal solution of the
problem. To assess the convergence speed to good approximation ratios we would
like an algorithm to approach values $\epsilon_{\psi} \approx 0$ in few
iterations. We also plot the frequency of sampling the ground state of the
problem Hamiltonian $\ket{\psi_\text{gs}}$:
\begin{equation}
    P_{\psi}(\text{gs}) = \vert \braket{\psi \mid \psi_\text{gs}}\vert^2.
\end{equation}
Ideally, we would like an algorithm to return the ground state with a frequency
$P_{\psi}(\text{gs}) \approx 1$, which implies small average energy
$\epsilon_\psi \approx 0$. The converse is not true because a superposition of
low-energy excited states $\ket{\psi}$ can exhibit a small average energy
$\epsilon_\psi \approx 0$ but small overlap with the ground state
$P_{\psi}(\text{gs}) \approx 0$ \cite{amaroFilteringVariationalQuantum2022}.

\subsection{Performance on 5-variable JSP}\label{ssec:comparison}

We analyzed all algorithms on a JSP instance with 5 free variables requiring 5
qubits. This is sufficiently small to run, essentially, on all available quantum
processors. We performed experiments for all VQA on a range of IBM quantum
processors. To make the results more comparable, all experiments in this section
use the same quantum processors, number of shots, ansatz (VQE, VarQITE, F-VQE)
and number of layers for each of the VQA (see Table~\ref{tab:VQA} for all
settings). We chose to highlight the results from the ibmq\_casablanca device in
the following plots since it showed the best final ground state frequency for
QAOA and good overall performance for VQE and VarQITE.
Appendix~\ref{sec:additional} presents additional hardware experiments for VQE,
QAOA and F-VQE and also VQE and QAOA results for CVaR quantile $\alpha=0.2$. The
goal of these experiments is to analyse the general convergence of the
algorithms without much fine-tuning and to select candidate algorithms for the
larger experiments in Sec.~\ref{ssec:scaling}.

\begin{figure}
    \includegraphics[width=4.23in]{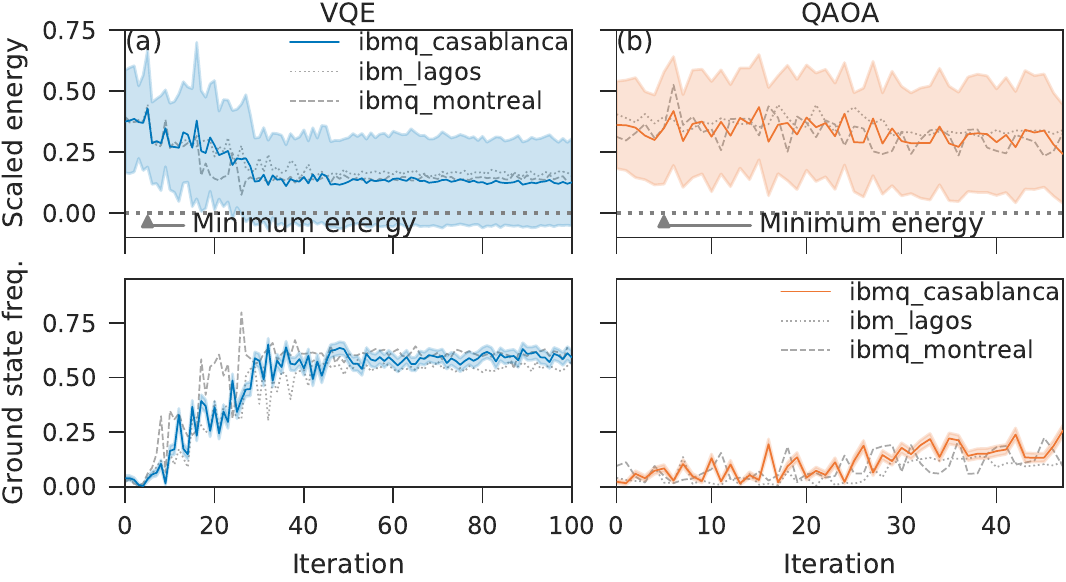}
    \caption{VQE and QAOA scaled energy $\epsilon_\psi$ (top panels) and ground
    state frequency $P_\psi(\text{gs})$ (bottom panels) for the JSP instance
    using 5 qubits and 1,000 shots on the IBM quantum processors indicated in
    the legend. The energy was rescaled with the minimum and maximum energy
    eigenvalues. Both VQA use the CVaR objective with $\alpha=0.5$. Error bands
    are the standard deviation (top panels) and 95\% confidence interval (bottom
    panels) (for clarity, error bands only shown for the solid line).}
    \label{fig:comparison_vqe_qaoa}
\end{figure}

\begin{figure}
    \includegraphics[width=4.23in]{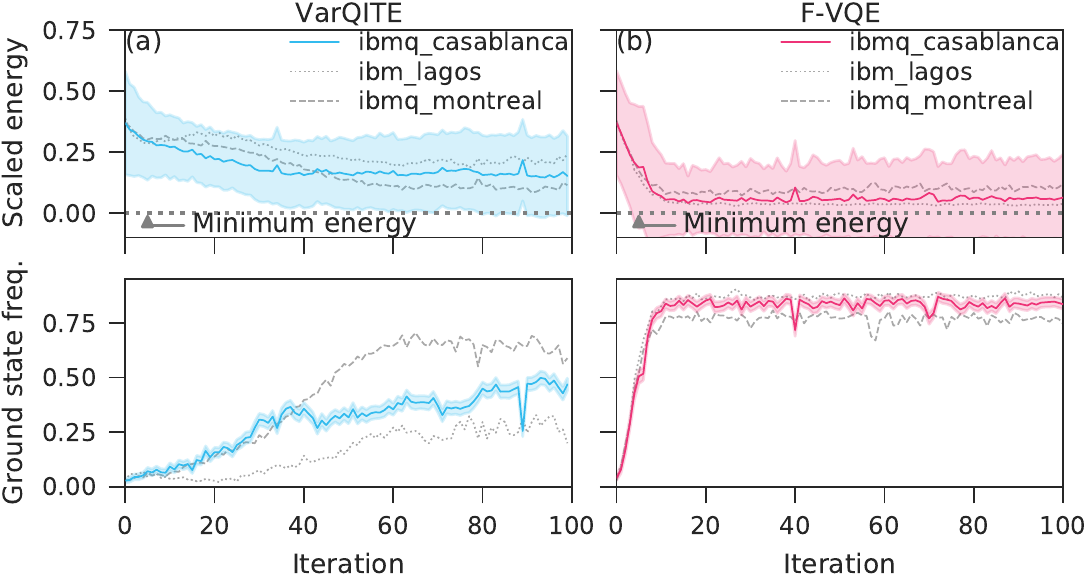}
    \caption{VarQITE and F-VQE scaled energy $\epsilon_\psi$ (top panels) and
    ground state frequency $P_\psi(\text{gs})$ (bottom panels) for the JSP
    instance using 5 qubits and 1,000 shots on the IBM quantum processors
    indicated in the legend. The energy was rescaled with the minimum and
    maximum energy eigenvalues. Error bands are the standard deviation (top
    panels) and 95\% confidence interval (bottom panels) (for clarity, error
    bands only shown for the solid line).}
    \label{fig:comparison_qite_fvqe}
\end{figure}

First, we analyzed VQE. Due to its simplicity it is ideal for initial
experimentation. We compared the CVaR objective with $\alpha < 1$ against the
standard VQE mean energy objective ($\alpha=1$). We observed that the CVaR
mainly leads to lower variance in the measurement outcomes.

Fig.~\ref{fig:comparison_vqe_qaoa}(a) shows the results for VQE using CVaR with
$\alpha=0.5$ and 1,000 shots and $p=2$ layers of the ansatz
Fig.~\ref{fig:ansatz}(a). VQE on ibmq\_casablanca converged after around 40
iterations with a frequency of sampling the ground state of approximately
59\%. The frequency of sampling the ground state is approximately bounded by the
value $\alpha$ of the CVaR. This is because CVaR optimises the left tail of the
empirical distribution up to quantile $\alpha$. If all the probability mass of
the distribution up to quantile $\alpha$ is on the ground state, the cost
function achieves its optimal value: the conditional expectation is the ground
state energy. At the same time, on average a fraction $1-\alpha$ of the
distribution sits in the right tail of excited states. Results for CVaR with
$\alpha=0.2$ in Fig.~\ref{fig:vqe-additional}(b) of App.~\ref{sec:additional}
are consistent with this observation. All quantum processors showed similar
final energies and ground state frequencies for VQE
(cf.~Fig.~\ref{fig:comparison_vqe_qaoa}(a)) with a moderate amount of variance
across devices during the initial iterations. Different choices of optimizers
could potentially improve convergence rate of
VQE~\cite{nanniciniPerformanceHybridQuantumclassical2019,stokesQuantumNaturalGradient2020}
but their fine-tuning was not in scope of this study.

QAOA with $p=2$ showed very slow convergence across all tested quantum
processors. The optimizer COBYLA terminated after 47, 50, 48 iterations for
ibmq\_casablanca, ibm\_lagos and ibmq\_montreal, respectively, when it was
unable to improve results further. Fig.~\ref{fig:comparison_vqe_qaoa}(b)shows
the scaled energy and ground state frequency with 1,000 shots and CVaR
$\alpha=0.5$ (same as VQE). In contrast to VQE, QAOA did not saturate the ground
state frequency bound at $\alpha$. We repeated QAOA experiments with CVaR
$\alpha=0.2$ on several quantum processors (see
Fig.~\ref{fig:qaoa-additional}(b)). In this case the ground state frequencies
saturated at around $\alpha=0.2$ but final average energies showed similar
performance as the $\alpha=0.5$ case.

Apart from the optimizer, a contributing factor of this poor performance is
likely that the QAOA ansatz is not hardware-efficient, i.e. the compiler needs
to add SWAP gates for routing. On ibmq\_casablanca the compiler embedded the
problem on qubits 1-3-4-5-6 (see Fig.~\ref{fig:ansatz}(b) for the device's
connectivity). In our instance each layer $p$ requires six 2-qubit operations of
the form $e^{-i \theta Z_iZ_j}$ each requiring 2 CNOTs. For $p=2$ layers this is
a total of 24 CNOTs to implement the unitaries $U(\gamma_1), U(\gamma_2)$.
Routing requires an additional 6 SWAPs, which are implemented with 3 CNOTs each,
for  a total of 18 CNOTs for routing. In total QAOA required 42 CNOTs. In
contrast, the hardware-efficient ansatz Fig.~\ref{fig:ansatz}(a) for the other
VQA can be embedded on a linear chain such as 0-1-3-5-6. This requires no SWAPs
and results in a total of 8 CNOTs for our VQE and F-VQE runs. The challenge of
scaling QAOA on quantum processors with restricted qubit connectivity was also
highlighted in \cite{harriganQuantumApproximateOptimization2021} and our results
appear to confirm that QAOA running on NISQ hardware requires fine-tuned
optimizers even for small-scale
instances~\cite{lavrijsenClassicalOptimizersNoisy2020,sungUsingModelsImprove2020}.

VarQITE converged somehwat more gradually compared to VQE but reached similar
final mean energies as VQE.  Figure~\ref{fig:comparison_qite_fvqe}(a) shows its
performance on different quantum processors with 1,000 shots and $p=2$ layers of
the ansatz Fig.~\ref{fig:ansatz}(a). In contrast to VQE, VarQITE exhibited a
higher variance of the final mean energy and ground state frequency across
different quantum processors. One of the issues of VarQITE is inversion of the
matrix $A$ in Eq.~\eqref{eq:optimizer-QITE}, which is estimated from measurement
shots. This can lead to unstable evolutions. Compared to QAOA, for our problem
instance VarQITE converged much faster and smoother across all quantum
processors.

F-VQE converged fastest on all quantum processors. Moreover,
Fig.~\ref{fig:comparison_qite_fvqe}(b) shows that its convergence is very
consistent across devices and the final mean energies are closest to the minimum
compared to the other VQA. F-VQE also showed high probability of sampling the
optimal solution after just 10-15 iterations, and high final probabilities of
84\%, 87\% and 75\% after 100 iterations on ibmq\_casablanca, ibm\_lagos and
ibmq\_montreal, respectively. We repeated the F-VQE experiment with a single
layer of an ansatz using a linear chain of CNOTs instead of the CNOT pattern in
Fig.~\ref{fig:ansatz}(a) with, essentially, identical results (not shown). This
confirms the fast convergence of this algorithm first observed for the weighted
MaxCut problem in Ref.~\cite{amaroFilteringVariationalQuantum2022}. Another
advantage of F-VQE compared to VarQITE is that F-VQE does not require inversion
of the---typically ill-conditioned---matrix $A$ in
Eq.~\eqref{eq:varqite_evolution}, which is estimated from measurement samples.
Based on these results we chose to focus on F-VQE for scaling up to larger JSP
instances.

\subsection{Performance on larger instances}\label{ssec:scaling}

\begin{figure}
    \includegraphics[width=4.23in]{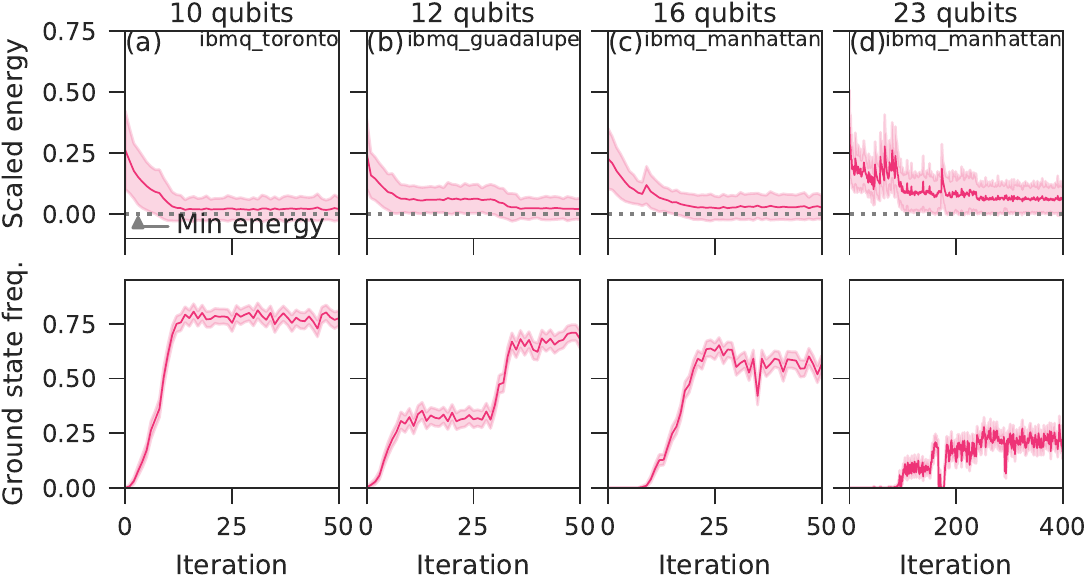}
    \caption{F-VQE scaled energy (top panels) and ground state frequency (bottom
    panels) for different JSP instances with (from left to right) $N=10$
    (ibmq\_toronto, 500 shots), $N=12$ (ibmq\_guadalupe, 550 shots), $N=16$
    (ibmq\_manhattan, 650 shots) and $N=23$ qubits (ibmq\_manhattan, 450 shots).
    The energy was rescaled with the maximum energy eigenvalue. Error bands are
    the standard deviation (top panels) and 95\% confidence interval (bottom
    panels).}
    \label{fig:comparison_fvqe}
\end{figure}

This section analyzes the effectiveness of F-VQE on larger JSP instances
executed on NISQ hardware. Figure~\ref{fig:comparison_fvqe} summarises the
results for up to 23 qubits executed on several IBM quantum processors. For
practical reasons (availability, queuing times on the largest device) we ran
those experiments on different processors.  However, based on the results in
Sec.~\ref{sec:results} we expect similar performance across different quantum
processors. F-VQE converges quickly in all cases. All experiments reach a
significant nonzero frequency of sampling the ground state: $P_{\psi}(\text{gs})
\approx 80\%$ for 10 qubits, $P_{\psi}(\text{gs}) \approx 70\%$ for 12 qubits,
$P_{\psi}(\text{gs}) \approx 60\%$ for 16 qubits, and $P_{\psi}(\text{gs})
\approx 25\%$ for 23 qubits.

An interesting case is $N=12$ (Fig.~\ref{fig:comparison_fvqe}(b)). From
iteration 10-30 F-VQE sampled the ground state and one particular excited state
with roughly equal probability. However, the algorithm was able to recover the
ground state with high probability from iteration 30.

The $N=23$ results show convergence in terms of the scaled energy and ground
state frequency. F-VQE sampled the ground state for the first time after 45
iterations and gradually builds up the probability of sampling it afterwards.
This means F-VQE is able to move to a parameter region with high probability of
sampling the optimal solution in a computational space of size $2^{23}$ despite
device errors and shot noise.

To our knowledge, the 23-qubit experiment is one of the largest experimental
demonstrations of VQA for combinatorial optimization. Otterbach~\emph{et al.}
\cite{otterbachUnsupervisedMachineLearning2017} demonstrated QAOA with $p=1$ on
Rigetti's 19-qubit transmon quantum processor.  Pagano~\emph{et al.}
\cite{paganoQuantumApproximateOptimization2020} demonstrated the convergence of
QAOA ($p=1$) for up to 20 qubits on a trapped-ion quantum processor. In
addition, they present QAOA performance close to optimal parameters with up to
40 qubits without performing the variational parameter optimization.
Harrigan~\emph{et al.} \cite{harriganQuantumApproximateOptimization2021}
demonstrated QAOA on Google's superconducting quantum processor Sycamore for up
to 23 qubits when the problem and hardware topologies match ($p=1, \dots, 5$)
and up to 22 qubits when the problem and hardware topologies differ ($p=1,
\dots, 3$). 

\section{Conclusions}\label{sec:conclusions}

In this case study, we solved a combinatorial optimization problem of wide
industrial relevance---job shop scheduling---on IBM's superconducting,
gate-based quantum processors. Our focus was on the performance of four
variational algorithms: the popular VQE and QAOA, as well as the more recent
VarQITE and F-VQE. Performance metrics were convergence speed in terms of the
number of iterations and the frequency of sampling the optimal solution. We
tested these genuinely quantum heuristics using up to 23 physical qubits.
   
In a first set of experiments we compared all algorithms on a JSP instance with
5 variables (qubits). F-VQE outperformed the other algorithms by all metrics.
VarQITE converged slower than F-VQE but was able to sample optimal solutions
with comparably high frequency. VQE converged slowly and sampled optimal
solutions less frequently. Lastly, QAOA struggled to converge owing to a
combination of deeper, more complex circuits and the optimizer choice. QAOA
convergence can possibly be improved with a fine-tuned
optimizer~\cite{sungUsingModelsImprove2020}. In the subsequent set of
experiments, we focused on F-VQE as the most promising algorithm and studied its
performance on increasingly large problem instances up to 23 variables (qubits).
To the best of our knowledge, this is amongst the largest combinatorial
optimization problems solved successfully by a variational algorithm on a
gate-based quantum processor.

One of the many challenges for variational quantum optimization heuristics is
solving larger and more realistic problem instances. It will be crucial to
improve convergence of heuristics using more qubits as commercial providers plan
a 2- to 4-fold increase of the qubit number on their flagship hardware in the
coming years.\footnote{See development roadmaps by IBM
\url{https://research.ibm.com/blog/ibm-quantum-roadmap} and Quantinuum (formerly
Honeywell Quantum Solutions)
\url{https://www.honeywell.com/us/en/news/2020/10/get-to-know-honeywell-s-latest-quantum-computer-system-model-h1},
for instance (accessed on 2022-02-04).} Our experiments suggest that F-VQE is a
step in this direction as it converged quickly even on the larger problems we
employed. Another challenge on superconducting quantum processors with hundreds
of qubits is sparse connectivity and cross-talk noise. F-VQE can address this
concern with ans\"atze that are independent of the problem's connectivity and
that can be embedded in a quantum processor's topology with lower or even zero
SWAP gate overhead from routing. In addition, error mitigation post processing
can be utilized~\cite{suguruHybridErrorMitigation2021}, although recent results
show that this requires careful analysis as these techniques can either improve
or hinder trainability of VQA~\cite{wangCanErrorMitigation2021}. Trapped-ion
quantum hardware may be soon equipped with dozens of qubits. Their low noise
levels and all-to-all qubit connectivity should be more suitable for deeper and
more complex ans\"atze. Hence, trapped-ion quantum processors may benefit from
the combination of F-VQE with \emph{causal
cones}~\cite{amaroFilteringVariationalQuantum2022}. Causal cones can split the
evaluation of the cost function into batches of circuits with fewer
qubits~\cite{benedettiHardwareefficientVariationalQuantum2021}. This allows
quantum computers to tackle combinatorial optimization problems with more
variables than their physical qubits and parallelize the workload.

The combination of the results of this case study together with the
aforementioned algorithmic and hardware improvements paint the optimistic
picture that near term quantum computers may be able to tackle combinatorial
optimization problems with hundreds of variables in the coming years.

\section{Abbreviations}

\begin{description}
    \item[COBYLA] Constrained Optimization By Linear Approximation
    \item[CVaR] Conditional Value-at-Risk
    \item[F-VQE] Filtering Variational Quantum Eigensolver
    \item[JSP] Job Shop Scheduling problem
    \item[NISQ] Noisy Intermediate-Scale Quantum
    \item[PQC] Parameterized Quantum Circuit
    \item[QAOA] Quantum Approximate Optimization Algorithm
    \item[QUBO] Quadratic Unconstrained Binary Optimization
    \item[VarQITE] Variational Quantum Imaginary Time Evolution
    \item[VQA] Variational Quantum Algorithms
    \item[VQE] Variational Quantum Eigensolver
\end{description}

\appendix

\section{Derivation of the QUBO formulation of the JSP}\label{sec:derivation}

This appendix describes the derivation of the QUBO formulation of the JSP in
Eq.~\eqref{eq:qubo}.

The cost of a schedule comprises three parts: the early delivery cost, late
delivery cost and production cost. The early and late delivery costs are a
penalty added when a job $j$ passes the last machine $M$ before or after its
\emph{due time} $d_j$, respectively:
\begin{equation}
	u_j(\bm{x}) = c_e \sum_{t=1}^{d_j} (d_j - t)x_{Mjt} + c_l \sum_{t=d_j+1}^{T_M} (t - d_j)x_{Mjt} \quad \forall\; j = 1, \dots, J.
\end{equation}
The constants $c_e$ and $c_l$ determine the magnitude of the early and late
delivery cost, respectively. Figure~\ref{fig:instance} illustrates the 20-job
instance used in our results together with its optimal schedule.

The production cost is a penalty added for production group switches of two jobs
entering a machine at subsequent time slots. The production group of job $j$ for
machine $m$ is determined by a matrix with entries $P_{mj}$. For each machine
$m$ we define a matrix $ G^{(m)}$ with entries
\begin{equation}
    G^{(m)}_{j_1j_2} = \begin{cases}
        0 & \text{if } P_{mj_1} = P_{mj_2},\\
        1 & \text{otherwise}.
    \end{cases}
\end{equation}
Hence, the production cost for machine $ m $ is given by
\begin{equation}
	s_m(\bm{x}) = c_p \sum_{j_1,j_2=1}^J\sum_{t=1}^{T_m-1} G^{(m)}_{j_1j_2} x_{mj_1t} x_{mj_2(t+1)} \quad \forall m=1, \dots, M.
\end{equation}
The constant $c_p$ determines the magnitude of the production cost.

The total cost of a schedule $\bm{x}$ is
\begin{equation} \label{eq:cost}
	c(\bm{x}) = \sum_{j=1}^{J} u_j(\bm{x}) +  \sum_{m=1}^{M} s_m(\bm{x}).
\end{equation}

We model the constraints of the JSP as additional cost functions.  The \emph{job
assignment constraints} enforces that each real job is assigned to exactly one
time slot in each machine
\begin{equation}\label{eq:real_job}
	g_{mj}(\bm{x}) \equiv \sum_{t=1}^{T_m} x_{mjt} = 1, \quad \forall m = 1, \dots, M\; \forall j = 1, \dots, J.
\end{equation}

The \emph{time assignment constraints} ensure that each time slot in each
machine is occupied by exactly one job:
\begin{equation} \label{eq:time}
	\ell_{mt}(\bm{x}, \bm{y}) = \left\{\begin{array}{ll}
		y_{mt} + \sum_{i=1}^{J} x_{mjt} &\text{for } 1 \leq t \leq e_m\\
		\sum_{j=1}^{J} x_{mjt} &\text{for } e_m < t \leq J\\
		1-y_{m(t-J)} + \sum_{j=1}^{J} x_{mjt} &\text{for } J < t \leq T_m
	\end{array}\right\} = 1
	\quad \begin{array}{ll}
	\forall t=1, \dots, T_m\\ \forall m=1, \dots, M.
	\end{array}
\end{equation}

The \emph{process order constraints} ensure that the processing time of a real
job does not decrease from one machine to the next:
\begin{equation} \label{eq:process_order}
	q_{mj}(\bm{x}) = \sum_{t=2}^{T_m} \sum_{t'=1}^{t-1} x_{mjt}x_{(m+1)jt'} = 0 \quad \forall m = 1, \dots, M-1\; \forall j = 1, \dots, J.
\end{equation}

The \emph{idle slot constraints} ensure that dummy jobs are placed before all
real jobs in each machine. Due to constraints $ \ell_{mt} $ in
Eq.~\eqref{eq:time} we only need to enforce that the transition from a real job
to a dummy job is prohibited at the beginning of a schedule:
\begin{equation} \label{eq:idle_slot_cons}
	r_{mt}(\bm{y}) = (1-y_{mt})y_{m(t+1)} = 0 \quad \forall t = 1, \dots, e_m - 1\; \forall m = 2, \dots, M.
\end{equation}
Note that constraints of this form are not required for machines with $e_m = 1$.

\section{Worst-case scaling of the JSP}\label{sec:jsp-scaling}
The total number of variables in the JSP formulation of Sec.~\ref{ssec:qubo} is
\begin{equation}
    N = \sum_{m=1}^M J (J + e_m) + e_m.
\end{equation}
The best-case scaling $\mathcal{O}(J^2M)$ is achieved for fixed $e_m$. In the worst case
the number of dummy jobs needs to grow by $J-1$ per machine to allow for a
complete reordering of all jobs. With the convention that $e_1=0$ this leads to
$e_m = (m-1)(J-1)$ and the worst-case scaling $\mathcal{O}(J^2 M^2)$.

Note that the dummy variables $ y_{m1} $ can be dropped from the problem for
every machine with $ e_m = 1 $. For $ e_m=1 $ the constraints $
\ell_{m1}(\bm{x}, \bm{y}) $ and $ \ell_{m(J+1)}(\bm{x}, \bm{y}) $ are
automatically satisfied given the rest of constraints for $ e_m=1 $. First, from
the rest of constraints $ \ell_{mt}(\bm{x}, \bm{y}) $ the $ J-1 $ time slots $ t
= 2, \dots, J$ contain one job. Second, from the constraints $g_{mj}(\bm{x})$
there are $ J $ jobs. Therefore, exactly one job is placed either in the time $
t=1 $ or in the time $ t=J+1 $ without the need of forcing the constraints $
\ell_{m1}(\bm{x}, \bm{y}) $ and $ \ell_{m(J+1)}(\bm{x}, \bm{y})$. 

It is possible to cut down the worst-case scaling to $\mathcal{O}(J^2M)$ with an
alternative formulation of the JSP. This alternative uses a binary encoding for
the $e_m$ dummy jobs. However, in this work we focused on fixed $e_m$ for all
instances, which leads to the same scaling. Furthermore, we fix most of the time
slots to the optimal solution and only leave the positions of a few jobs free.
This way we can systematically increase problem sizes and analyse scaling of the
algorithms.

\section{Quantum hardware}\label{sec:hardware}

Table~\ref{tab:ibm-devices} lists the quantum processors used in this work and
some of their basic properties at the time of execution. More information is
availale at \url{https://quantum-computing.ibm.com/services}.

\begin{table}[ht]
    \begin{tabular}{lccl}
        \hline
        Device & No. of qubits & Quantum volume  & Connectivity\\
        \hline
        ibmq\_belem & 5 & 16 & T-shaped\\
        ibmq\_casablanca & 7 & 32 & Fig.~\ref{fig:ansatz}(b)\\
        ibmq\_guadalupe & 16 & 32 & Heavy-hexagon~\cite{chamberlandTopologicalSubsystemCodes2020}\\
        ibmq\_jakarta & 7 & 16 & Fig.~\ref{fig:ansatz}(b)\\
        ibm\_lagos & 7 & 32 & Fig.~\ref{fig:ansatz}(b)\\
        ibmq\_manhattan & 65 & 32 &
        Heavy-hexagon\\
        ibmq\_montreal & 27 & 128 & Heavy-hexagon\\
        ibmq\_sidney & 27 & 32 & Heavy-hexagon\\
        ibmq\_toronto & 27 & 32 & Heavy-hexagon\\
        \hline
    \end{tabular}
    \caption{Hardware devices used in this study.}
    \label{tab:ibm-devices}
\end{table}

\section{Additional experiments}\label{sec:additional}
\begin{figure}
    \centering
    \includegraphics{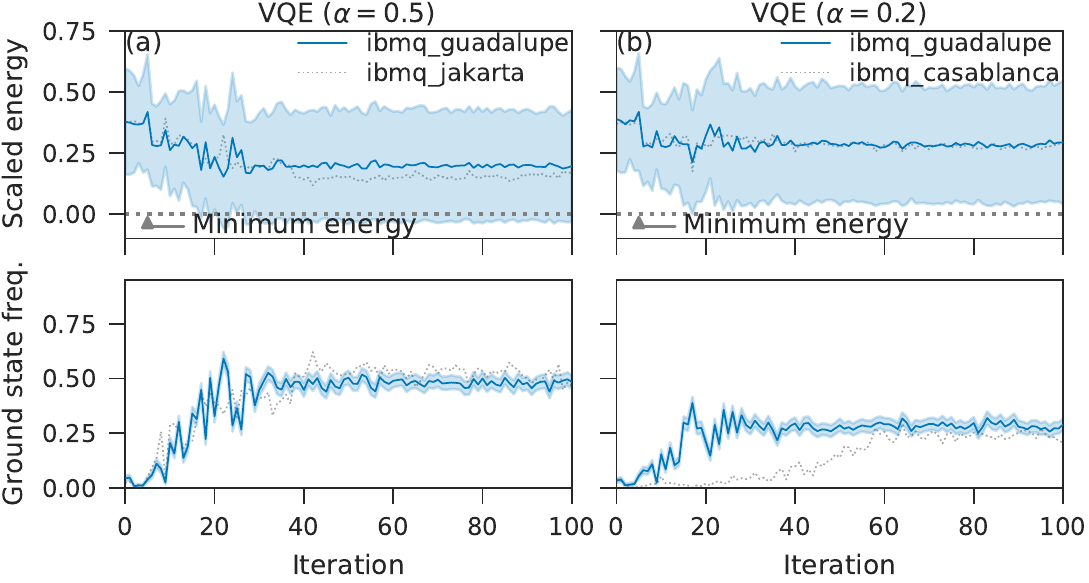}
    \caption{VQE scaled energy $\epsilon_\psi$ (top panels) and ground state
    frequency $P_\psi(\text{gs})$ (bottom panels) for the 5-qubit JSP instance
    with CVaR (a) $\alpha=0.5$ and (b) $\alpha=0.2$. For other settings, see
    Tab.~\ref{tab:VQA}. The energy was rescaled with the minimum and maximum
    energy eigenvalues. Error bands are the standard deviation (top panels) and
    95\% confidence interval (bottom panels) (for clarity, error bands only shown for the
    solid line).}
    \label{fig:vqe-additional}
\end{figure}

Figure~\ref{fig:vqe-additional} shows results of additional hardware experiments
for VQE with CVaR quantiles $\alpha=0.5$ (Fig.~\ref{fig:vqe-additional}(a)) and
$\alpha=0.2$ (Fig.~\ref{fig:vqe-additional}(b)) for the 5-qubit JSP instance
discussed in Sec.~\ref{ssec:comparison}. For all other parameters see
Tab.~\ref{tab:VQA}. In both cases VQE reaches a ground state frequency of
approximately $\alpha$ indicating that the CVaR objective was achieved.
Generally, the $\alpha=0.2$ case converged to a mean energy considerably further
from the optimal value than for $\alpha=0.5$.

\begin{figure}
    \centering
    \includegraphics{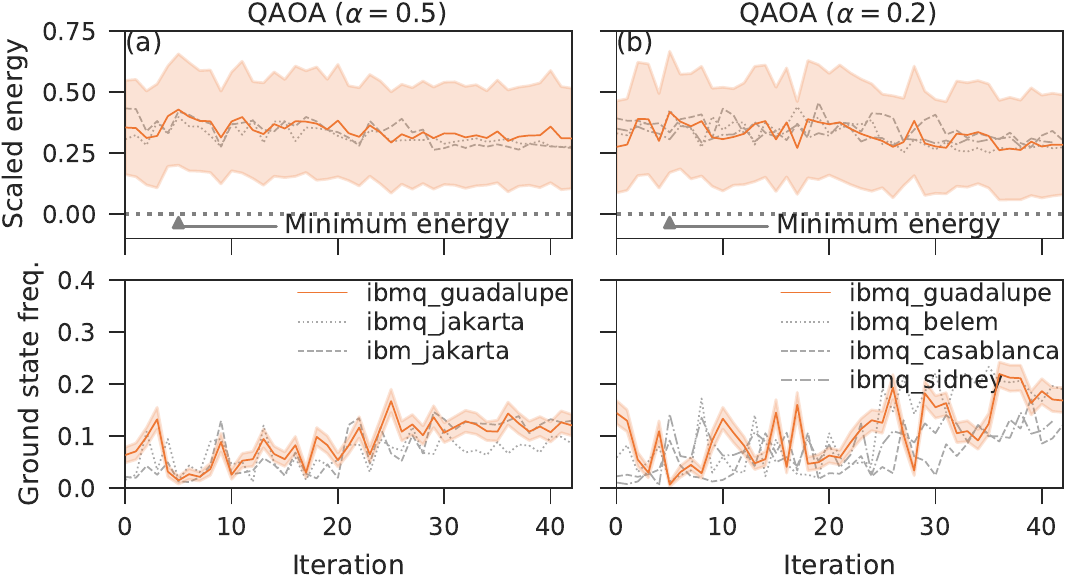}
    \caption{QAOA scaled energy $\epsilon_\psi$ (top panels) and ground state
    frequency $P_\psi(\text{gs})$ (bottom panels) for the 5-qubit JSP instance
    with CVaR (a) $\alpha=0.5$ and (b) $\alpha=0.2$. For other settings, see
    Tab.~\ref{tab:VQA}. The energy was rescaled with the minimum and maximum
    energy eigenvalues. Error bands are the standard deviation (top panels) and
    95\% confidence interval (bottom panels) (for clarity, error bands only
    shown for the solid line).}
    \label{fig:qaoa-additional}
\end{figure}

Figure~\ref{fig:qaoa-additional} shows results of additional hardware
experiments for QAOA with CVaR quantiles $\alpha=0.5$
(Fig.~\ref{fig:qaoa-additional}(a)) and $\alpha=0.2$
(Fig.~\ref{fig:qaoa-additional}(b)) for the 5-qubit JSP instance discussed in
Sec.~\ref{ssec:comparison}. For all other parameters see Tab.~\ref{tab:VQA}.
QAOA with $\alpha=0.2$ reaches a ground state frequency of approximately
$\alpha$ indicating that the CVaR objective was achieved in this case.

\begin{figure}
    \centering
    \includegraphics{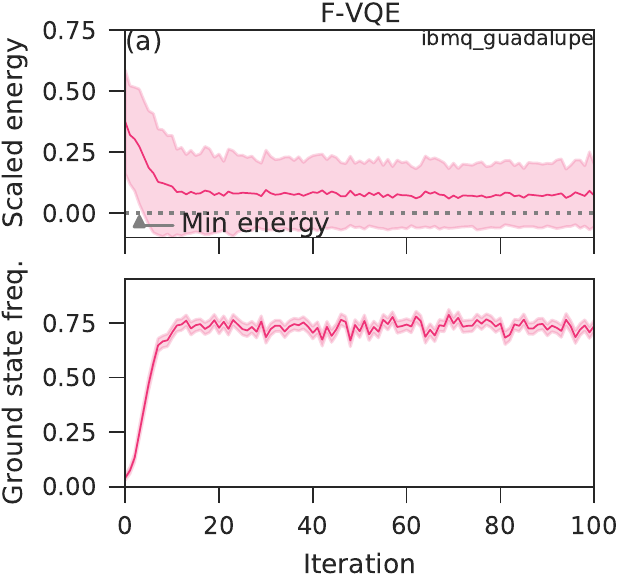}
    \caption{F-VQE scaled energy $\epsilon_\psi$ (top panels) and ground state
    frequency $P_\psi(\text{gs})$ (bottom panels) for the 5-qubit JSP instance.
    For other settings, see Tab.~\ref{tab:VQA}. The energy was rescaled with the
    minimum and maximum energy eigenvalues. Error bands are the standard
    deviation (top panels) and 95\% confidence interval (bottom panels).}
    \label{fig:fvqe-additional}
\end{figure}

Figure~\ref{fig:fvqe-additional} shows results of one additional hardware
experiment for F-VQE on ibmq\_guadalupe for the 5-qubit JSP instance discussed
in Sec.~\ref{ssec:comparison}. For all other parameters see Tab.~\ref{tab:VQA}.
The overall performance is comparable to its performance on other quantum
processors in Sec.~\ref{ssec:comparison}.

\section*{Availability of data and materials}
The datasets used and/or analysed during the current study are available from
the corresponding author on reasonable request.
    
\section*{Author's contributions}
All authors contributed to the drafting of the manuscript. DA, MF and MR
designed the work and experiments. KH designed the JSP instance analyzed in this
work. DA acquired the data. DA, MF and MR interpreted and analysed the data. DA
and NF created the software for this work. All authors read and approved the
final manuscript.

\section*{Acknowledgements}
The authors would like to thank Carlo Modica for helping with the execution of
some experiments on quantum hardware, and Michael Lubasch and Marcello Benedetti
for the helpful discussions.

\bibliography{references}

\end{document}